\documentclass{ws-procs11x85}
\usepackage{multicol}

\begin{document}


\vspace{2cm}
\title{Probing Universality Of Gravity}
\author{Naresh Dadhich}  
\address {Inter-University Centre for Astronomy \& Astrophysics,
Post Bag 4, Pune~411~007, India \\ E-mail: nkd@iucaa.ernet.in}

\maketitle

\abstract{I wish to expound a novel perspective of probing universal character of  gravity. To begin with, inclusion of zero mass particle in mechanics leads to special relativity while its interaction 
with a universal force shared by all particles leads to general relativity. 
The universal nature of force further suggests that it is intrinsically 
attractive, self interactive and higher dimensional. I argue that the 
principle of universality could serve as a good guide for future directions.}
\vskip2pc 



\begin{multicols}{2}

 The moment we admit the existence of zero mass particle, we need a new mechanics for it moves with the same universal speed relative to all observers. Newtonian mechanics cannot accommodate a universal speed. The new mechanics that emerges is the special relativity which synthesizes space and time into one whole 
- spacetime. Alongside come synthesis of energy, momentum and mass, and the light  cone structure of spacetime events. It is the principle of universality that has led us to SR. 

 A universal entity is the one which is freely accessible and shared by all particles. Space and time are without question the most primary universal 
entities we know of. 
There cannot in principle exist two unconnected universal things. How is one universal entity is distinguished from the other universal entity? This distinguishing feature will contradict the universal nature of them. Thus all 
universal things must be related and this relation would have also to be universal. It is hence pertinent to ask for a relation between the two universal entities; space and time. 
That will 
require a universal speed which is what is precisely provided by the zero mass particle - photon, a quantum of light. 

It is geometry that makes universal statement. For instance, 
motion of free particles is a straight line is a universal property which is 
 expressed as a geometric statement. A universal relation or property must therefore be expressible as a geometrical statement. The universality of speed of light must also be a geometric statement and the Minkowski geometry of 
spacetime is that statement. From this we wish to propound a general 
guiding principle, {\it any universal physical property/force must be 
expressible as a property of universal entities, space and time, through a 
geometric statement}.

What we have done is that in conformity with universality we have asked for 
incorporation of massless particles in mechanics, which is not admissible in  
the Newtonian mechanics. Taking into account the characteristic property of motion of massless 
particles we are forced to enlarge the Newtonian framework to Minkowski 
framework which admits both massive and massless particles. In our further 
exploration, this would be the guiding finger: {\it ask a question which 
is not admitted in the existing framework, then enlarge the framework as 
indicated/suggested by the question itself such that the question is answered.}

Of the four basic forces in Nature, gravitation distinguishes itself from 
the rest by the property of universality. As argued above, universal force must be unique. Let us hence begin by considering a universal force, and then 
show that it can be nothing other than gravity. Since the force is universal, 
it must be long range and  act on all particles including massless ones as well. But 
massless particle propagates with universal constant speed which cannot be changed. Then how do 
we make it feel the universal force, and feel it must? We thus face the 
contradiction that massless particle must feel the force but its speed must 
not change. This is certainly not possible in the existing framework. How do we enlarge 
the framework? Two suggestions come forward from our guiding principle, one 
since the force is universal it must be expressible as a geometric property of spacetime
 and two, what do we really want massless particle to do is that when it is skirting the source of the universal force 
it should acknowledge its presence by bending rather than going 
straight. To illustrate this point, let 
us consider a piece of wood floating in a river. It floats freely and bends 
as the river bends. No force really acts on it to bend its course, it simply 
follows the flow of the river. This suggests that bend the {\it river} in 
which massless particle propagates freely. Thus we arrive at the profound 
insight into the nature of the universal force that it must bend/curve space - 
rather spacetime in which massive as well as massless particles propagate 
freely. That is the envisaged universal force can only be described by 
curvature of spacetime and 
in no other way. Isn't it a remarkable and insightful conclusion that follows 
from the very simple common sense arguments? 

Thus it is the  universality of force which demands spacetime to respond to 
it by being its {\it potential}. It can no longer remain as inert 
background but instead it now has all the dynamics of the universal force. This is the unique and 
distinguishing property of universality. It then ceases to be an external force and its 
dynamics has to be fully governed by the spacetime curvature. This 
automatically incorporates the first suggestion mentioned above that 
universal force should become a property of spacetime. The equation of motion 
for the force does not have to be prescribed but must rather follow from the 
spacetime curvature all by itself. The curvature of spacetime is given by the 
Riemann curvature 
tensor for the metric $g_{ab}$ and it satisfies the Bianchi 
differential identity. The contraction of 
which yields the second rank symmetric tensor, constructed from the Ricci 
tensor, and it is divergence free. That is, 
\begin{equation} \label{div}
\nabla_{b}G^{ab} = 0 
\end{equation}
where 
\begin{equation}
G_{ab} = R_{ab} - \frac{1}{2}Rg_{ab}\,.
\end{equation}
The above equation implies 
\begin{equation} \label{gr}
G_{ab} = -\kappa T_{ab} - \Lambda g_{ab}
\end{equation}
with 
\begin{equation}
\nabla_{b}T^{ab} = 0
\end{equation}
where $T_{ab}$ is a symmetric tensor, and $\kappa$ and $\Lambda$ are 
constants. On the left we have a differential expression involving the second 
derivative and square of the first derivative of the metric which now acts 
as a potential for the universal force. Thus on the right should be the 
source of the force. What should be the source for a universal force? 
Something which is shared by all particles - matter energy. That identifies 
$T_{ab}$ with the energy momentum tensor of matter distribution and vanishing 
of its divergence ensures conservation of energy and momentum. In the weak 
field and slow motion limit, this equation for small $\Lambda$ takes the form 
$\nabla^2 \phi = {1\over2} \kappa \rho$ where $\rho$ is energy density of matter distribution. Hence the force in this limit is inverse square law. Further 
since the force is self interacting which is 
indicated by the presence of  square of first derivative of the metric in the Riemann curvature, it must have charge. What should be its polarity? We all know that for 
stability it is essential that total charge is zero. The matter energy is 
positive and hence charge neutrality will require field energy to have opposite charge;i.e. it be negative. Since the potential is negative for a long range force, it must hence be attractive - yet another remarkable result from simple 
common sense arguments. We thus have a inverse square law attractive universal force which must agree with the Newtonian gravitational law in the first approximation. That determines $\kappa = 8 \pi G/c^2$. The above equation is nothing but the Einstein equation for gravitation with a new constant $\Lambda$. The universal force we began with could be nothing other than gravitation and thus gravity is the unique universal interaction \cite{n1}. The universality of force also tells us that all its properties are 
self determined and we have no freedom to put any constraints on it. For 
instance the Einstein equation is valid in all dimensions where Riemann 
curvature can be defined ($D\ge2$). {\it The dimensionality of space would have to be determined by the dynamics of gravitational field itself.}   

 Note that the constant $\Lambda$ enters into the equation naturally and is 
in fact at the same footing as the energy momentum tensor. It all follows from the Riemann curvature which contains the whole dynamics. It appears as a constant of integration in the field equation, and its Newtonian analogue would be 
$\nabla^2 \phi = k = const.$  This gives rise to radially symmetric  harmonic 
oscillator potential. It is addition of a constant to the field equation which has come about because now dynamics of the field cannot be prescribed but it 
follows from the spacetime geometry. It is the geometry which brings in this 
new constant which is a true new constant of the 
Einstein gravity. What value should it have has to be determined by observation/experiment? It makes a very important 
statement that even when space is free of all non-gravitational matter energy 
distribution, empty space has non-trivial dynamics. The above equation refers 
to spacetime in entirety which means whatever can be done in it has to be  
included in it. It is well known that vacuum can suffer quantum 
fluctuations which produce stress energy tensor precisely of the form 
$\Lambda g_{ab}$ and it must be 
included in the equation. Had Einstein followed this chain of arguments, he 
would have anticipated 
gravitational effect of quantum fluctuations of vacuum and would have made a 
profound prediction rather than a profound blunder. 

Classically a constant scalar field with a constant potential has no dynamics while in GR it again generates 
precisely the stress tensor of the type $\Lambda g_{ab}$ due to constant potential. That of course has 
non trivial dynamics. In other perspective we can thus say that  $\Lambda$ is the measure of the constant potential to which vacuum has been raised.\footnote{In the Schwarzschild field, we have the Newtonian potential, 
$\phi = -M/r$, note that a non-zero constant, which is classically inert, 
cannot be added to it.
The Einstein equation determines the potential of an isolated body absolutely 
with its zero fixed at infinity and nowhere else \cite{n3}. That is for the 
local situation and here we have constant potential in a global setting.} 
 That means setting $\Lambda = 0$ is to put vacuum to absolute zero potential. Note that  spacetime responds to gravity by curving, and it also mediates 
motion of massless particles. It should therefore possess micro-structure 
which can facilitate bending (curvature) due to gravity and propagation of 
massless particles as a property of spacetime. It could then naturally suffer quantum fluctuations. That means it would thus be natural for vacuum to have 
the inherent dynamics of $\Lambda g_{ab}$. The 
big question however is what value should $\Lambda$ have? The answer to this 
question is that its value has to be determined empirically by experiment 
that probes the micro-structure of spacetime. 

Let us now turn to dimensionality of the gravitational interaction. Since 
the equation is valid for any $D$, it is therefore inherently higher 
dimensional interaction. However let us go by the principle of minimum requirement. For the full realization of dynamics of gravity, the minimum dimension 
required is $4=3+1$. Let us keep time aside, so we require $3$-space which could be inhibited by its source, matter energy distribution. Would gravity be 
confined to $3$-space/brane or could it propagate off the brane? To establish 
the attractive character of the universal force, we had earlier argued that 
the field energy should have charge and its polarity be negative (opposite 
of positive polarity of matter energy). This was invoked to have overall 
charge neutrality. Here negative charge is spread all over the space and 
hence complete charge neutrality could only be achieved when integration is 
taken all over the space. So long as the total charge is non-zero on any finite surface, then the field must propagate off the surface. For instance, if 
net electric charge on a sphere is non-zero, the electric field does 
propagate off the sphere into $3$-space. Similarly since total gravitational 
charge in a finite region of space cannot be fully neutralized,
the field will have to propagate off the $3$-brane into ($5=4+1$)-D bulk 
spacetime. However as we include larger and larger region on the brane, charge 
strength diminishes to zero 
asymptotically. This is perhaps suggestive of the fact that gravity cannot go 
deep enough into the bulk but rather remains confined near to the brane. If 
the matter fields are confined to 
$n$-space, the field will leak into the $(n+1)$th dimension but will not 
propagate deep enough. That is, massless graviton will have ground state on 
the brane and hence will remain confined to the brane. This is exactly what 
is required 
to happen for the brane world gravity \cite{add,rs}. It is remarkable that we 
have here motivated the higher dimensional and the 
brane world nature of the gravitational field purely from classical 
standpoint. In particular we 
can have the Randall - Sundrum brane world model \cite{rs}. In that case, 
the confinement of gravity on the brane requires that bulk spacetime must be 
anti-deSitter (AdS) with negative $\Lambda$ in the bulk. It has been shown 
for the AdS bulk and flat brane system that the Newtonian gravity can be 
recovered on the brane with high energy $1/r^3$ correction to the potential 
\cite{rs,gt}. The most interesting case is of Schwarzschild - AdS bulk 
harbouring FRW brane. In that case localization of gravity on the FRW brane 
requires non-negative effective $\Lambda$ on the brane \cite{rk,pd}. 

The universality of gravitation demands that field is self interactive and  
spin 2 non-Abelian field. Self interaction is however an iterative process 
and the Riemann curvature includes the first iteration through square of 
first derivative of the metric. The Riemann curvature led to the Einstein 
equation. How about the next order of iteration involving higher powers of 
the derivative of the metric? That is inclusion of higher powers of the 
Riemann curvature. At any rate, the resulting equation should be second 
order quasilinear (i.e. second derivative must occur linearly). Then the 
question is, what is the most general second order quasilinear equation which 
can be derived from polynomial in Riemann curvature. It turns out that there 
exists the specific Gauss - Bonnet (GB) combination involving the square of 
Riemann, Ricci and scalar curvature (and there is the Lovelock 
generalization for higher order polynomial, see for instance \cite{d1,d2}) 
which still yields the 
second order quasilinear equation for the field. However the GB 
term is topological for $D<5$;i.e. it makes non-trivial contribution only for 
$D>4$. That is higher order iteration effects of self interaction will 
therefore be non-trivial only in the higher dimensions $D>4$. The high energy 
gravity would naturally involve higher order iterative GB term which can 
have non-trivial meaning only in higher dimensions. This is therefore an 
independent new argument for higher dimensionality of gravity. 

Now if we adhere to our minimum requirement principle, the minimum 
number of spatial dimension required for full realization of gravitational 
dynamics is $3$ and hence there is no compelling reason for non-gravitational 
matter energy to exist in dimension greater than $3$. Gravity could however 
leak into the fourth (bulk) spatial dimension. In the $5$-D bulk, the GB term 
attains dynamics which accounts for higher order self interaction effects and 
hence must be included. It turns out that dS or AdS automatically solves the GB term in the equation and it simply redefines the bulk $\Lambda$. It suggests that the dynamics of the bulk spacetime may be governed by the GB 
term;i.e. the source for the AdS bulk in the Randall - Sundrum brane 
world model may fully or partially be provided by the GB term \cite{sami}! As 
a matter of fact the GB term will generate effective negative $\Lambda$ and 
thereby AdS in the bulk naturally because it is the measure of gravitational 
field energy in higher order iteration. As argued elsewhere \cite{n2}, the 
positive energy condition for the field energy is that it be negative. 

It is however well known that charge neutrality is a necessary condition for 
stable equilibrium but not sufficient. The Earnshaw's theorem states that it 
is impossible to attain stable equilibrium purely under 
electromagnetic force. This is because the field has two kinds of charges 
which are isolated and localized. On the other hand in the case of gravity 
the other (negative) charge is distributed all over the space and hence is 
not isolated and localizable. A gravitational situation could be envisioned as 
follows: A positive charge (body) sitting in its own field which has negative 
charge spread around it in space like a net. This system is obviously stable. 
It is the distributed nature of the other charge that provides the stability. 
It is thus no surprise that systems bound by gravity are always stable. 

Let us for a moment digress to quantum theory. What question should we ask 
which can lead us from classical to quantum mechanics? We have two kinds of 
motion, particle and wave. Like particle wave must also carry energy and 
momentum with it. So it must like particle have a $4$-momentum vector while 
on the other hand its motion is completely determined by the $4$-wave vector. 
Since both these vectors refer to the same wave, they must be proportional. 
This gives the basic quantum mechanical relations between energy and 
frequency, and momentum and $3$-(wave)vector. From this it is easy to get to 
the uncertainty and commutation relations which form the basic quantum 
principle. 

First of all we have not yet succeeded in writing the uncertainty principle 
in a spacetime covariant form. Further the quantum principle is universal. 
Going by the guiding principle of 
universality, it must, like the speed of light, be expressible as a property 
of spacetime. This has unfortunately not happened. What is required is 
exactly what Minkowski did to SR by synthesizing the speed of light into 
the spacetime structure. This is however very difficult because synthesis of 
quantum principle with the spacetime would ask for discrete structure which 
is in contradiction with the inherent continuum of spacetime. However so long 
as this doesn't happen, quantum theory will remain incomplete. Thus for 
completion of quantum theory it would be required that spacetime must have 
micro-structure which could accord to quantum principle. It is the geometry 
of that which would synthesize quantum principle with the spacetime. This 
is an open question of over 100 years standing. 

The same question is also coming up from the gravity side as well. Unlike 
quantum theory, GR is complete like classical electrodynamics. However 
at high energy we have quantum electrodynamics. We know that at high energy 
matter attains quantum character,i.e. $T_{ab}$ on the right becomes quantum. 
On the left is the spacetime which would now have to become quantum - 
discrete. This is what being pursued in the canonical approach of loop quantum gravity \cite{lqg}. It is a programme of quantizing gravitational field; i.e. quantizing spacetime in $4$ dimensions. On the other hand string theory begins 
with the quantum principle and SR to construct a consistent theory of matter. 
That naturally leads first to $26$ and then to $10$ or $11$ dimensions. 
Gravity is considered as a spin $2$ gauge field in higher dimensional flat 
spacetime. There is however no unique way of getting down to the usual $4$ 
dimensions. It makes connection with GR as a low energy effective theory 
\cite{gw}. The two approaches appear to be 
complimentary, either catching some aspect of the problem. The strong point 
of the former is spacetime background independent formulation while that of 
the latter is the gauge field framework which is shared by other fields and 
thereby its strong orientation to unification of the fields. However 
asymptotically the two would have to converge when the complete 
theory comes about. 

Both gravity at high energy as well as completion of quantum theory require 
discrete micro-structure for spacetime. This suggests that there may 
perhaps be one and the same answer to both the questions. In a sense the two 
approaches, string theory and loop quantum gravity anchor respectively to 
quantum field theoretic and geometric gravitational aspects. 

Adhering to our guiding principle of universality, what question should we 
ask and how should we enlarge the existing framework to answer the question? 
Since spacetime is the fundamental universal entity, let us ask do there 
still remain some properties of it which have yet remained untapped? One is 
dimensionality of space and the other is its (non) commutativity. The former 
is however essential for the string theory and is also quite in vogue in the 
brane world models. We have however attempted to articulate a new and simple 
minded motivation purely based on classical consideration for higher 
dimensions. On the other hand non commutativity 
of space has also attracted some attention 
recently and it is hoped that it might facilitate in building up discrete  
structure for space. 

 Returning to the enigmatic $\Lambda$, we would like to argue that it is 
really anchored on the micro-structure of spacetime and hence might hold the 
ultimate key to the problem. It may in a deep sense connect through some 
duality relation micro with macro structure of the Universe. It defines a new 
scale given 
by the Einstein gravity. On the other hand we already have the Planck 
length which is not given by any theory but constructed by using three 
universal constants. I believe that it is not a wise thing to take the Planck 
length a priori fundamental but instead we should attempt to deduce it in a 
fundamental manner. $\Lambda$ on the other hand is a scale provided by a 
fundamental theory and hence should be respected. We thus have two length 
scales where we require only one and hence it is natural to expect 
that there should exist a relation between them which will perhaps encode a 
profound physical truth. 
 
Let me come back to the guiding question one should ask? In 
the gravitational field equation, we have the curvature of spacetime on the 
left, and matter stress tensor and the vacuum energy $\Lambda$ on the right. 
If $T_{ab}$ lives only in $3$-brane which becomes quantum at high 
energy, while vacuum energy can still support a continuum bulk spacetime. 
Taking the cue from what we have discussed so far what question should we ask 
and how should we enlarge the framework to admit and answer the question? The 
universality demands that the Einstein equation should remain valid whether 
the matter field source is classical or quantum. For the quantum case, 
the spacetime curvature will be required to have discrete quantum character. 
Should that mean that spacetime itself becomes quantum? This is what is being 
explored by the canonical loop quantum gravity approach \cite{lqg}.
 In this approach we remain bound to the $4$-dimensional spacetime and there 
is no background spacetime relative to which spacetime is being quantized. In 
the string theory approach, we are already in higher dimensions and gravity 
is being considered as massless spin 2 field and GR appears as the low energy 
effective theory in $4$ dimensions along with plethora of other fields 
\cite{gw}. In either case it is adaption of radically new framework which 
cannot be seen as enlargement of the existing framework.

One possible enlargement of the framework could be: for matter fields 
confined to $3$-brane, the $5$-dimensional bulk spacetime with $\Lambda$ and 
GB support could provide the continuum background for gravity 
(spacetime curvature) to be quantized 
on the $3$-brane. This is the suggestion that crops up in the spirit of what 
has been done in going from Newton to Einstein. However the pertinent 
question is whether it is technically and conceptually workable? That has to 
be investigated. This suggestion is in the spirit of the guiding principle we 
have propounded. Perhaps it may lead to some insight. 

Let me reiterate that we have here enunciated a method of asking question 
which is motivated by the principle of universality, and the question also 
suggests enlargement of the framework such that it gets answered. And we 
arrive at a new framework. One of the remarkably interesting applications of 
this method is to establish why the universal force has to be attractive? 
This is perhaps the simplest and most direct demonstration of why gravity is 
attractive. Equally illuminating is the physical 
realization of the higher order iteration of self interaction 
through the GB contribution, and thereby the higher dimensional nature 
of gravity. Following this train of thought what we need to do is to ask the 
right kind of question which will show us the way beyond GR or quantum 
theory. The main problem is to identify the right question. 

We have argued that one of the most challenging problems is to understand 
$\Lambda$ in terms of the basic building blocks of spacetime. Any 
quantization scheme should have to address to it \cite{gp}. We believe   
that there must exist some fundamental relation between $\Lambda$ and the 
Planck length which needs to be discovered (recently there have come up a 
couple of such proposals \cite{paddy}). That may hold the key to the 
problem. 

These are some of the rumbling thoughts which I wanted to share \cite{n4}.

\section*{Acknowledgement}
 
I wish to thank warmly the organizers of the Regional 
conference for their kind invitation and wonderful hospitality which gave me 
the opportunity to share and propound a new perspective. I also thank 
Sudhendu Rai Choudhury for helpful discussions.


\end{multicols}

\end{document}